\documentclass[galaxies,review,accept,moreauthors,pdftex,10pt,a4paper]{mdpi} 

\setitemize{parsep=6pt,itemsep=0pt,leftmargin=*,labelsep=5.5mm}
\setenumerate{parsep=6pt,itemsep=0pt,leftmargin=*,labelsep=5.5mm}
\setlist[description]{itemsep=0mm}  
\usepackage{upgreek}
\firstpage{1} 
\pubvolume{7}
\issuenum{2}
\articlenumber{49}
\pubyear{2019}
\copyrightyear{2019}
\history{Received: 1 March 2019; Accepted: 17 April 2019; Published: 24 April 2019}
\updates{yes} % If there is an update available, un-comment this line
\usepackage{amsfonts,amssymb,amsmath,graphicx,hyperref}
\Title{The Important Role of Cosmic-Ray Re-Acceleration}

\Author{Martina Cardillo\orcidA{}}%Please carefully check the accuracy of names and affiliations. Changes will not be possible after proofreading.

\AuthorNames{Martina Cardillo}

\address[1]{%
Istituto Nazionale di Astrofisica-Istitituto di Astrofisica e Planetologia Spaziali, Via del Fosso del Cavaliere 100, 00133 Roma, Italy; martina.cardillo@inaf.it; Tel.: +39-06-4993-4462}

\abstract{In the last decades, the improvement of high energy instruments has enabled a deeper understanding of the Cosmic Ray origin issue. In particular, the $\gamma$-ray satellites AGILE (Astrorivelatore Gamma ad Immagini LEggero) and Fermi-LAT (Fermi-Large Area Telescope) have strongly contributed to the confirmation of direct involvement of Supernova Remnants in Cosmic Ray energization. Despite several attempts to fit experimental data assuming the presence of freshly accelerated particles, the scientific community is now aware that the role of pre-existing Cosmic Ray re-acceleration cannot be neglected. In this work, we highlight the importance of pre-existing Cosmic Ray re-acceleration in the Galaxy showing its fundamental contribution in middle aged Supernova Remnant shocks and in the forward shock of stellar winds.}


\begin{document}

%%%%%%%%%%%%%%%%%%%%%%%%%%%%%%%%%%%%%%%%%%
\section{Introduction}
\label{Sec:intro}
%%%%%%%%%%%%%%%%%%%%%%%%%%%%%%%%%%%%%%%%%%%
Cosmic Ray (CR) particles are mainly protons ($90\%$) and nuclei with a spectrum that extends over 15 orders of magnitude in energy ($E = [10^{10} \div 10^{20}]$ eV). Such a large energy range can be explained with different kinds of sources, both galactic and extragalactic. CR origin is one of the most discussed topics of high energy astrophysics, with~$\gamma$-ray astronomy playing a critical role in this field of research. Indeed, neutral pions $\pi^{0}$, produced by CR proton-target proton interaction decay in two $\gamma$-rays with the same spectral behavior of parent population but with lower energies and fluxes. Particle and $\gamma$-ray spectra revealed several tricky characteristics that questioned theoretical models and leave open very important~questions.

The first issue is CR spectral composition at $E\sim3\times10^{15}$ eV (the so called \textit{knee}), where the spectra steepens probably because of the transition from CR light component (hydrogen, helium and CNO) to CR heavy component (mainly Iron) (\cite{Blasi13, Amato14} for a review). However, in~the last years, ARGO (Astrophysical Radiation with Ground-based Observatory)~\cite{Demitri14} and YAC1-TIBET Array (Yangbajing Air shower Core TIBET array)~\cite{Huang13} measurements showed a steepening of the light component spectrum at 650~TeV, well below the PeV \textit{knee} detected by KASCADE-Grande (KArlsruhe Shower Core and Array DEtector Grande)~\cite{Apel13} and all previous instruments. If, on the~one hand, ARGO results could solve the complex Pevatron problem of CR Galactic component, on~the other hand, it makes the understanding of transition energy location harder~\cite{Amato14, Cardillo15}.
%Please define all acronyms here.

Another important issue was introduced by PAMELA (Payload for Antimatter Matter Exploration and Light-nuclei Astrophysics) and AMS-2 (Alpha Magnetic Spectrometer) satellite measurements. Their data revealed a hardening of proton and helium spectra above a rigidity of 200 GV, and~a rising in the positron fraction in an energy range from a few GeV to 100 GeV~\citep{Adriani09,Aguilar13,Adriani11,Aguilar14, Aguilar15a,Aguilar15b}. Several theoretical models were developed to explain these spectral features, taking into account different possible explanations: differential diffusion and turbulence~\citep{Neronov12, Kachelriess12, Tomassetti12, Blasi12, Aloisio13} or possible contribution from local sources~\citep{Thoudam12,Thoudam13} for spectral hardening, and~dark matter~\citep{Fan10} or Pulsar Wind Nebula (PWN)~\citep{Bucciantini11} for positron fraction rising (\citep{Blasi13,Amato14} and reference therein). However, a~final explanation is still~missing.

From the $\gamma$-ray point of view, AGILE~\citep{Tavani09} and Fermi-LAT~\citep{Atwood09} satellites, together with large Cherenkov telescopes HESS (High Energy Stereoscopic System)~\citep{Hinton04} MAGIC (Major Atmospheric Gamma-ray Imaging Cherenkov Telescope)~\citep{Ferenc05} and VERITAS (Very Energetic Radiation Imaging Telescope Array System)~\citep{Krennrich04}, have collected a large amount of data from young ($t_{age}<1000$ years) and middle-aged ($t_{age}>1000$ years) Supernova Remnants (SNRs). 
These data provided direct evidence of CR energization in SNR shocks~\citep{Giuliani11,Ackermann13,Cardillo14,Morlino12} but no direct proof of freshly accelerated particle presence. Moreover, detected $\gamma$-ray spectra showed some peculiarities, such as energy spectral index steeper than 2, deviating from linear and Non Linear Diffusive Shock Acceleration (DSA and NLDSA) theory predictions~\citep{Blasi13, Amato14}. In~spite of the several models developed to explain these features, which introduced ``ad~hoc'' broken power-law distribution, low-energy cut-off (e.g., \citep{Giuliani11,Ackermann13}), or~turbulence damping~\citep{Malkov11}, we are still looking for a more consistent interpretation of experimental~data.

Until a few years ago, contribution from re-acceleration of pre-existing CRs was totally neglected because it was considered sub-dominant with respect to a possible particle acceleration at the source. However, some recent works showed how stochastic re-acceleration in Galactic turbulence~\citep{Osborne88} could effectively contribute to the overall CR spectrum~\citep{Drury15,Drury17} or explain $\gamma$-ray emission from Fermi-Bubbles~\citep{Cheng15}. Moreover, theoretical models~\citep{Uchiyama12,Lee15,Tang15,Cardillo16,Blasi17} and very recent numerical simulations~\citep{Caprioli18} proved that the Diffusive Shock Re-Acceleration (DSRA) at the source, like in an SNR shock, can give a dominant contribution to CR particle and $\gamma$-ray spectra. Here we point out the importance that DSRA could have to understand the observed $gamma$-ray spectra. Indeed, DSRA of pre-existing CRs can well explain the $\gamma$-ray emission from two different galactic sources, the~SNR W44~\citep{Cardillo16} and the OB-star $\kappa$-Ori stellar wind~\citep{Cardillo19}.

This brief review is focused only on the Galactic component of the CR spectrum. In~Section~\ref{Sec:reacceleration}, we summarize the most important features of stochastic and diffusive particle re-acceleration. In~Section~\ref{Sec:model} we describe an analytical model used to model experimental data from $\gamma$-ray sources. In~Section~\ref{Sec:examples} we briefly show results obtained in the case of SNR W44 and $\kappa$-Ori star wind. Then we write our conclusions in Section~\ref{Sec:Conclusions}.

%%%%%%%%%%%%%%%%%%%%%%%%%%%%%%%%%%%%%%%%%%
\section{The Re-Acceleration~Contribution}
\label{Sec:reacceleration}
%%%%%%%%%%%%%%%%%%%%%%%%%%%%%%%%%%%%%%%%%%

CR particles fill our Galaxy almost isotropically. During~their propagation, they interact with an interstellar magnetic field and gain energy through scattering on random magnetic perturbations with scale of order of their Larmor radius. This process is a second order Fermi acceleration ~\citep{Fermi49}, also called stochastic re-acceleration, and~could explain some CR particle spectrum features, such as spectral hardening or electron-positron fraction rising (see Section~\ref{Subsec:stochastic}). 

However, pre-existing CRs could be re-accelerated not only during their propagation but also in correspondence of local sources (e.g., SNR shocks). Indeed, CR re-acceleration at the source, as is the case for the pure acceleration, is a first order Fermi energization mechanism, which is more efficient than stochastic re-acceleration. This could explain the overall particle CR spectrum features where stochastic re-acceleration fails (see Section~\ref{Subsec:DSRA}). Moreover, in~recent years, analysis of $\gamma$-ray spectra from some middle-aged SNRs pointed out some peculiar features not easily explained by typical DSA models but in agreement with re-acceleration at the~source.  

%%%%%%%%%%%%%%%%%%%%%%%%%%%%%%%%%%%%%%%%%%
\subsection{The LIS Spectrum from Voyager~1}
\label{Subsec:Voyager}
%%%%%%%%%%%%%%%%%%%%%%%%%%%%%%%%%%%%%%%%%%

Because of solar modulation, the~first direct measurement of a low-energy CR very Local Inter Stellar (LIS) spectrum was in 2013 when the Voyager 1 spacecraft entered in the heliopause, where modulation effects are negligible~\citep{Webber13, Potgieter13}. New data collected by Voyager I strongly contributes to the most important results obtained in a re-acceleration~context.

A parametrization of CR-sea in our Galaxy was done in~\cite{Potgieter13,Potgieter14, Bisschoff15}, for~protons, helium and electrons, separately~\citep{Cardillo16}. The spectrum of protons and He nuclei can be described as:
%*********************
\begin{equation}
J_{LIS,n}=A_h\, \left(\frac{E^{a}}{\beta_{p}^{2}}\right)\left(\frac{E^{d}+k^{d}}{1+k^{d}}\right)^{-b}\ ,
\label{Eq:LIS_p}
\end{equation}
%********************
where $E$ is particle kinetic energy per nucleon in units of $[GeV/n]$, and~$J_{LIS,n}$ is in units of $[particles/m^{2}/s/sr/(GeV/n)]$. Parameter values in Equation~(\ref{Eq:LIS_p}) are: $a=1.03$, $b=3.18$, $d=1.21$, $k=0.77$, $A_h=3719$ for proton LIS, $J_{LIS,p}$; and $a=1.02$, $b=3.15$, $d=1.19$, $k=0.60$, $A_h=195.4$ for helium LIS, $J_{LIS,He}$.

The LIS electron flux, instead, can be described as
%***************
\begin{equation}
J_{LIS,e}= 0.21\left(\frac{E^{1.35}}{\beta_{e}^{2}}\right)\left(\frac{E^{1.65}+0.6920}{1.6920}\right)^{-1.1515}+\left(1.73\, \exp\left[4.19-5.40\, \ln\,(E)-8.9\,E^{-0.64}\right]\right)
\label{Eq:LIS_e}
\end{equation}
%*********************
in units of $[{ particles}/{ m}^{2}/{ s}/{ sr}/{ MeV}]$~\citep{Potgieter13}. Here $E$ is the electron kinetic energy expressed in units of GeV, whereas the last term fits the PAMELA (and AMS-02) data in the $[5-20]$ GeV~range.

Data from Voyager I, together with data at higher energies ($E\geq1$ GeV/n) provided by PAMELA~\citep{Adriani11} and AMS02~\citep{Aguilar14, Aguilar15a, Aguilar15b} satellites, give us a complete description of pre-existing CR spectrum, critical to develop realistic re-acceleration~models.

%%%%%%%%%%%%%%%%%%%%%%%%%%%%%%%%%%%%%%%%%%
\subsection{Stochastic~Re-Acceleration}
\label{Subsec:stochastic}
%%%%%%%%%%%%%%%%%%%%%%%%%%%%%%%%%%%%%%%%%%

CRs propagate in our Galaxy through scattering by resonant magnetic perturbations, moving with Alfv\'en velocities $V_{A}=\frac{B_0}{\sqrt(4\pi\rho_{ISM})}$, which is strictly correlated with the static average magnetic field, $B_{0}$, and~the medium density, $\rho_{ISM}$. During~propagation, CRs gain energy through stochastic re-acceleration~\citep{Fermi49, Thornbury14}. This physical process was introduced in~\cite{Osborne88} to explain low-energy secondary to primary ratio decrement with energy. They found that re-acceleration due to perturbations with a Kolmogorov spectrum (scaling with $k^{-5/3}$ where $k$ is the wave number) could provide the behavior shown by experimental data. About ten years later, in~\cite{Heinbach95} the authors confirmed that result, proving how diffusive re-acceleration model was in good agreement with data without introducing further escape~mechanisms. 

In the last years, in~\cite{Thornbury14} and~\cite{Drury15} the authors gave a qualitative estimation of transferred power from inter-stellar turbulence damping to CRs. Assuming a power-law momentum distribution with a spectral index $\alpha= 4.8$ (as provided by PAMELA and AMS-02~\citep{Adriani09, Aguilar13}), they found that most of the power comes from the low-energy part of CR spectrum and that re-accelerated particles contribute with about 20$\%$ to the global CR Galactic~spectrum.

A quantitative estimation of the transferred power was done in~\cite{Drury17}, taking into account very LIS spectrum measured by Voyager I. They develop both an analytical and a numerical (using the package GALPROP~\citep{Strong98}) model that can fit new measurements of boron over carbon (B/C) ratio from AMS-02 satellite~\citep{Aguilar16}, confirming a Kolmogorov-like turbulence spectrum. Moreover, they found that energy input obtained through re-acceleration mechanism spans from about $25\%$ to $50\%$ of the total energy. This result confirms that most of the energy is provided by sub-relativistic particles and implies that interstellar turbulence has to be damped by CRs to transfer them its energy. In~order to explain time and physical scale able to provide this high energy budget, the~author of~\cite{Blasi17} considered the possibility of a contribution from re-acceleration at the source (see the next Section~\ref{Subsec:DSRA}).

Stochastic re-acceleration could even explain $\gamma$-ray emission from Fermi-Bubbles~\citep{Dobler10, Su10} as shown in the work of~\cite{Cheng15}. This extended $\gamma$-ray emission above and below the Galactic Center implies a large number of high energy particles at about 10 kpc from Galactic disk. Numerical simulations pointed out that electrons from SNR shocks penetrating the bubbles cannot provide the electron density required to explain the observed $\gamma$-ray flux and the microwave spectrum detected by the Planck satellite~\citep{Ade13}. Hence, the~authors developed a model that considers both stochastic re-acceleration and convective transport inside the bubble due to the effect of Galactic wind~\citep{Breitschwerdt91, Breitschwerdt02, Bloemen93}. The~combined effect of these two mechanisms is more efficient (about one order of magnitude) than the acceleration and can explain both $\gamma$-ray and microwave data. In~\cite{Mertsch18} the authors confirmed this scenario, revealing that multi-wavelength spectral and morphological behavior of Fermi Bubbles could be explained by IC and synchrotron emission from electrons re-accelerated by Rayleigh-Taylor-like turbulence generated in a supersonic outward flowing shell. 
%The authors computed the electron density  Numerical simulations revealed that electrons from SNR shocks penetrating the bubbles cannot explain this emission. Consequently, they introduced a propagation model via diffusion, and consequent re-acceleration, inside the bubbles. By adding in CR transport equation the effect of Galactic wind  and adiabatic losses, they fit. Then, in~\cite{Mertsch18} the authors have implemented particle kinetic equation, describing acceleration and transport inside the bubbles. They have taken into account both stochastic acceleration by turbulence and low-Mach number shock in correspondence of bubble boundaries, where a Rayleigh-Taylor-like instability may arise. They found that multi-wavelength spectral and morphological behavior of Fermi Bubbles could be explained by IC and synchrotron emission from electrons accelerated by turbulence generated in a supersonic outward flowing shell. However, a final accepted explanation is still missing.

Based on these results, we conclude that stochastic re-acceleration is fundamental in the understanding of CR behavior and the same is valid for diffusive shock re-acceleration, as~we will see in the next~section.

%%%%%%%%%%%%%%%%%%%%%%%%%%%%%%%%%%%%%%%%%%
\subsection{Diffusive Shock Re-Acceleration at the~Source}
\label{Subsec:DSRA}
%%%%%%%%%%%%%%%%%%%%%%%%%%%%%%%%%%%%%%%%%%

After the first detection of $\gamma$-ray emission below 200 MeV from a SNR, SNR W44~\citep{Giuliani11, Ackermann13, Cardillo14}, followed by similar detections in other middle aged SNRs, like IC443~\citep{Ackermann13} and W51c~\citep{Jogler16}, the~whole CR community was excited by the chance to have an unambiguous evidence of CR acceleration in a SNR shock. Indeed, direct evidence of CR acceleration in a source is possible in the $\gamma$-ray band where we can distinguish the electronic component to the CR hadronic one. From~one hand, a~detection of 100~TeV photons would imply hadronic origin of the emission because the electronic Inverse Compton (IC) is sub-dominant at that energy due to Klein-Nishina cross section suppression. On~the other hand, a detection of the ``pion bump'' signature at about $70$ MeV, due to neutral pion rest mass ($\sim$135~MeV), could also allow us to distinguish hadronic contribution from the leptonic Bremsstrahlung one. However, all $\gamma$-ray detected middle-aged SNRs not only have a cut-off at very low-energy but also have $\gamma$-ray spectra with features difficult to understand considering linear and non-linear DSA theory. They revealed a very steep high energy index, $\alpha\geq3$, that can be fitted only by adding an ``ad~hoc'' low-energy cut-off at a parent power-law distribution~\citep{Giuliani11} or using a broken power-law distribution~\citep{Ackermann13, Cardillo14}. Moreover, middle-aged SNRs have a slow shock velocity ($v_{s}\sim~$10$^{2}$ km/s) that could hardly explain efficient CR~acceleration.

The possibility of pre-existing CR re-acceleration contribution at the W44 shock was introduced by~\citep{Uchiyama10} in order to explain the steep spectrum detected by Fermi-LAT~\citep{Abdo10}. This model takes into account the possible formation of a thin compressed adiabatic shell due to Molecular Cloud (MC)/SNR interaction that enhances $\gamma$-ray emissivity after re-acceleration (the ``crushed-cloud'' model from~\cite{Blandford82}). In~\cite{Lee15}, the~authors developed an analogous model analyzing even its temporal evolution. Both works found similar results: re-accelerated pre-existing CRs can explain the SNR W44 spectral behavior. However, there was an open issue: in both models, ``ad~hoc'' spectral features are necessary to obtain a good fit of experimental~data.

In 2016, Voyager 1 data outside the heliosphere allowed to consider the very LIS spectrum of pre-existing CRs. This spectrum was taken into account in~\cite{Cardillo16} where the authors developed a re-acceleration model with a thin compressed shell but also introducing some important aspects: helium nuclei from Voyager data; radio emission only in the region of AGILE detected $\gamma$-ray emission and not in the whole radio extension of the remnant; a simple PL distribution with high-energy cut-off. This model is briefly described in Section~\ref{Sec:model} and the results obtained for W44 in Section~\ref{Subsec:W44}. 

The re-acceleration process could explain $\gamma$-ray emission not only from other SNRs similar to W44, as~IC443, W51c and W49b, but~also from different kinds of objects. At~the end of December 2018, the~AGILE satellite has confirmed a Fermi-LAT detected residual emission~\citep{Ackermann12} from a region around the OB star $\kappa$-Ori~\citep{Marchili18}, in~correspondence of a CO-detected star forming region. A~re-acceleration model applied at the star wind forward shock (FS) could explain this emission as described forward in Section~\ref{Subsec:Orion}~\citep{Cardillo19}.

DSRA was then considered in~\cite{Blasi17} to explain some features of the overall CR spectrum, because~of the very stringent conditions required by stochastic re-acceleration~\citep{Drury17}. In~that work, the~author investigates the hardening, detected by PAMELA and AMS02~\citep{Adriani09, Aguilar15a,Aguilar15b, Yan17}, of~protons, helium nuclei and other heavier elements,  validating his model with new collected data on B/C ratio~\citep{Aguilar16}. He showed that re-acceleration models can explain primary (Carbon and Oxygen) spectra, regardless of diffusion coefficient changes and without further amount of grammage, as~provided by other models~\cite{Aloisio15}. However, the~author highlights that re-acceleration cannot explain the enhancement of positron fraction~\citep{Amato14}. This conclusion is important because points out that re-acceleration process must be considered in all CR models but it cannot replace the main contribution from freshly accelerated~particles.

Finally, recent numerical simulations developed by~\cite{Caprioli18} prove that pre-existing CRs with sufficiently high energy can be reflected at the shock. This mechanism generates turbulence through streaming instability (both resonant~\cite{Lagage83} and non-resonant~\cite{Bell04}), affecting shock inclination. In~this way, even quasi-perpendicular shocks (with an angle $\geq 70^{\circ}$) can efficiently accelerate thermal particles, despite of the results obtained in simulations without re-acceleration~\citep{Caprioli15}. Moreover, in~determined and physically plausible conditions, CR seed current is constant, regardless of shock Mach number, shock inclination and CR velocities. A~quantitative estimation of re-acceleration effect in SNRs provides an efficiency of a few percent, that can be higher in presence of high density target, in~perfect agreement with the W44-like~cases.

In the following, we will focus on the two different cases of $\gamma$-ray emission seen above, introducing some fundamental aspects of the model of re-acceleration at the source used to fit~data.

%%%%%%%%%%%%%%%%%%%%%%%%%%%%%%%%%%%%%%%%%%%
\section{A DSRA~Model}
\label{Sec:model}
\unskip
%%%%%%%%%%%%%%%%%%%%%%%%%%%%%%%%%%%%%%%%%%
\vspace{-6pt}
    %**************************************************
    \subsection{Re-Acceleration and~Acceleration}
    \label{Subsec:Re+Ac}
    %**************************************************

Particle re-acceleration at the shock of a source is a Fermi first order acceleration process of pre-existing CRs, without~a ``real'' thermal injection~\citep{Caprioli14}. The~critical condition is that, in~the upstream infinity, particle distribution equals the Galactic one, $f(x=-\infty,p)=f_{\infty}(p)$. Solving CR transport equation with this requirement~\citep{Blasi04, Cardillo16}, we obtain reaccelerated particle spectral distribution:
%---------------------------------------------------
\begin{equation}
f_{0}(p)=\alpha\left(\frac{p}{p_{m}}\right)^{-\alpha}\int^{p}_{p_{m}}\frac{dp'}{p'}\left(\frac{p'}{p_{m}}\right)^{\alpha}f_{\infty}(p')
\label{Eq:Re}
\end{equation}
%------------------------------------------------------------------------
with $\alpha= \frac{3r_{sh}}{r_{sh}-1}$, where $r_{sh}=\frac{u_{d}}{u_{u}}$ is the compression ratio at the shock and $u_{d}$ and $u_{u}$ are downstream and upstream velocities, respectively. $p_{m}$ is the minimum momentum in Galactic CR spectrum.  $f_{infty}(p)$ is the Galactic CR spectrum obtained from the fluxes described in Section~\ref{Subsec:Voyager}, using the usual expression $4 \pi p^{2} f_{\infty,i} (p) dp = \frac{4\pi}{v(p)} J_{LIS,i} (E) dE$. 

The main effects of re-acceleration can be summarized as~follows:
\begin{itemize}
 \item enhancement of particle momentum up to a maximum value dependent on time scales of the system (SNR age, acceleration time, energy losses);
  \item hardening of parent spectra steeper than $p^{-\alpha}$ 
\end{itemize}

Freshly accelerated particles could be present even if re-acceleration is the dominant process. They can be described as a simple power-law particle distribution, according to the DSA model:
%-----------------------------------
\begin{equation}
f_{i}(p)=k_{i}\left(\frac{p}{p_{inj}}\right)^{-\alpha}
\label{Eq:acc}
\end{equation}
%-----------------------------------
where the index $i$ indicates protons or electrons and $p_{inj}$ is the injection momentum corresponding to an injection energy $E_{inj}\sim4.5E_{sh}$, with~$E_{sh}=\frac{1}{2}m_{p}v_{sh}^{2}$~\citep{Caprioli14}. The normalization value for protons, $k_{p}$, is computed from balance between CR pressure, $P_{CR},$ and ram pressure of the shock, $\xi_{CR}\rho_{0}v_{sh}^{2}$~\citep{Cardillo16}. $\rho_0$ is upstream density in the assumption of a totally ionized medium at the shock, and~$\xi_{CR}$ is CR acceleration efficiency. Normalization of electron distribution, $k_{e}$, is then fixed by assuming the most conservative CR electron/proton ratio, $k_{ep}\approx10^{-2}$. Normalization strongly depends on target medium density and on shock velocity. Consequently, in~middle aged SNRs like W44, with~a slow shock velocity ($\sim$100 km/s), the~chance to have freshly accelerated particles is very low. Acceleration can enhance normalization of total $\gamma$-ray spectrum and, if~dominant, it modifies its~shape.

    %**************************************************
    \subsection{Crushed Cloud~Presence}
    \label{Subsec:crushed}
    %**************************************************
    
Since most of SNRs detected in $\gamma$-ray band are interacting with a MC, the~effect of a SNR/MC interaction on CR spectrum has to be considered. In~1982, the~authors of~\cite{Blandford82} introduced the ``crushed cloud'' model: if the pre-shock medium density is sufficiently high (as in a MC), a~thin compressed radiative shell forms behind the shock, ionizing all its propagation region~\citep{Cardillo16}. Shell compression can be stopped by magnetic or thermal pressure and we can obtain an estimation for compressed density, $n_m$, from~their balance with shock ram pressure:
%---------------------------------------------------
\begin{displaymath}
n_{0}\mu_{H}v_{sh}^{2}=\left\{ \begin{array}{ll}
\frac{B_{m}^{2}}{8\pi} & \textrm{Magnetic pressure;}\\
n_{m}K_{B}T& \textrm{Thermal Pressure}
\end{array} \right. 
\end{displaymath}
%--------------------------------------------------
where $B_{m}=\sqrt{\frac{2}{3}}\left(\frac{n_{m}}{n_{0}}\right)B_{0}$ is compressed magnetic field, $\mu_{H}$ is the mass per hydrogen nucleus,\linebreak $B_{0}=b\sqrt{n_{0}/cm^{-3}}$ $\upmu$G is unperturbed magnetic field upstream of the shock, where $b$ depends on Alfv\'en velocity $V_{a}$ and can vary in the range [0.3--3] in MCs~\citep{Hollenbach89,Crutcher99}. $K_{B}$ is the Boltzmann constant and $T$ is shell temperature. According to the approach described in~\cite{Draine11}, compressed shell temperature is stable around $10^{4}$ K, because~cooling efficiency drops abruptly for higher~temperatures.%Please check if this should be a multiplication or endush

Computation of compressed density is critical because strictly correlated with spectral changes due to adiabatic compression. Indeed, due to this further compression, the~final spectrum will be:
%---------------------------------------
\begin{equation}
f'(p)=f_{0}(s^{-1/3}p)\ ,
\label{Eq:compressed spectrum}
\end{equation}
%---------------------------------------
where $s$ is the compression factor is equal to
%---------------------------------------
\begin{equation}
s\equiv \left(\frac{n_{m}}{n_{d}}\right)=\left(\frac{n_{m}}{r_{sh} n_{0}}\right)
\label{Eq:factor s}
\end{equation}
%----------------------------------------
where $n_{d}$ is density immediately downstream of the shock, $n_{m}/n_d$ is compression ratio due to radiative cooling and $r_{sh}=n_{d}/n_{0}$ is the shock compression ratio due to the shock that we have already~seen. 

The whole scenario can be summarized as~follows:
\begin{itemize}
 \item Galactic Cosmic Rays are re-accelerated by first order Fermi energization mechanism in the interaction region between shock and its environment. Their spectrum, if~steeper than the slope $\alpha$ due to shock compression ratio, will become hard as $\alpha$. Initial density of the upstream medium, $n_{0}$, is compressed of a factor $r_{sh}$;
 \item in the right conditions, freshly accelerated particles can be injected at the shock through the first order Fermi mechanism;
 \item with sufficiently high density, a~thin adiabatic shell forms behind the shock. There, energized particles undergo to a further compression before escape, with~a consequent enhancement of their energy. The~density is enhanced by a further factor $s$.
\end{itemize}

    %**************************************************
    \subsection{Energy~Losses}
    \label{Subsec:losses}
    %**************************************************

Finally we have to consider energy losses. Protons lose energy through pp-interaction~\citep{Kelner06} and ionization, at~high and low energies, respectively. Electrons are affected by synchrotron, Inverse Compton and Bremsstrahlung losses~\citep{Ginzburg69} at the highest energies, and~by ionization at the lowest energies. The~kinetic equation becomes:
%------------------------------------------
\begin{equation}
\frac{\partial N_{i}(E,t)}{\partial t}=\frac{\partial}{\partial E}\left[b(E)N_{i}(E,t)\right]+Q_{i}(E),
\label{Eq:kinetic}
\end{equation}
%------------------------------------
where $b(E)=-\frac{dE}{dt}$ takes into account all losses and $Q_{i}(E)$ is particle injection rate per unit energy interval, derived from energized and compressed spectrum (see \citep{Cardillo16} for details).

The fundamental condition for CR energization is that acceleration time, $t_{acc}$, is lower than the lowest value between the MC/SNR interaction time, $t_{int}$, and~the loss time, $t_{loss}$. The~acceleration time can be expressed as $t_{acc}\approx D(p)/v_{sh}^{2}$~\citep{Blasi13, Amato14}, where $D(E)=\frac{1}{3}r_{L}c\left(\frac{L_c}{r_L}\right)^{k_T - 1}$ \cite{Ptuskin03} is the diffusion coefficient, $r_{L}$ is particle Larmor radius, $L_{v}$ is perturbation correlation length and $k_T$ is perturbation spectral index~\citep{Cardillo16}. From~the condition $t_{acc}< min(t_{int},t_{loss})$, CR distribution have a cut-off at a maximum momentum equal to
%----------------------------
\begin{equation}
p_{\rm max}\propto \left(B_{0}\right)\left(v_{sh}\right)^{\frac{2}{1-\delta}}\left(t_{min}\right)^{\frac{1}{1-\delta}}\left(L_{c}\right)^{-\frac{\delta}{1-\delta}}\ .
\label{Eq:pmax_gen}
\end{equation}
%----------------------
where magnetic power spectrum index, $\delta=k_{T}-1$, depends on turbulence model considered~\citep{Cardillo16}. 

    %**************************************************
    \subsection{$\gamma$-Ray~Emission}
    \label{Subsec:gray_emission}
    %**************************************************
    
Detected $\gamma$-ray emission is produced by CR protons and electrons energized and compressed, able to overcome energy losses. Protons contribute through pp-interactions, producing neutrals and charged pions that, in~turn, produce $\gamma$-rays and secondary electrons. The~produced fluxes can be described following the formalism of~\cite{Kelner06}:
%****************
\begin{displaymath}
\Phi_{s}(E_s)=\left\{ \begin{array}{ll}
2\times\int^{\infty}_{E_{min}}\frac{f_{\pi}(E_{\pi})}{\sqrt{E_{\pi}^{2}-(m_{\pi}c^2)^{2}}}dE_{\pi} & \textrm{Low energies.}\\
v\,n_{0}\int^{1}_{0}\sigma_{inel}(E_s/x)f(E_s/x)F_s(x,E_s/x)\frac{dx}{x}   & \textrm{High energies;}
\end{array}\right.
\end{displaymath}
%***************************
where $s$ refers to secondary species considered, either electrons or $\gamma$-rays, $x=E_s/E_{n}$, where $E_{n}$ is the nucleon energy, $F_s$ is an analytical function describing spectral distribution of secondaries, $v$ is nucleon velocity and $f(E_s/x)=f(E_n)$ is our final particle distribution, that considers both protons and helium contributions. Inelastic cross section, $\sigma_{inel}$, depends on particle energy and on its ratio with energy threshold for $\pi^{0}$ production. At~lower energies, we used $\delta$-approximation, where the factor 2 accounts for generation of two photons from every neutral pion and both $e^+$ and $e^-$ in the case of charged pions. $E_{min}$ is the pion minimum energy necessary to produce a secondary particle with energy $E_s$, and~$f_{\pi}$ is the production rate of pions with energy $E_\pi$ (see \citep{Kelner06, Cardillo16} for details).

Now we summarize how this model could explain $\gamma$-ray and radio emission from the SNR W44 and $\kappa$-Ori stellar~wind.

%%%%%%%%%%%%%%%%%%%%%%%%%%%%%%%%%%%%%%%%%%
\section{Two Important Cases: A Supernova Remnants and an OB Star~Wind}
\label{Sec:examples}
%%%%%%%%%%%%%%%%%%%%%%%%%%%%%%%%%%%%%%%%%%%%%

For W44, DSRA was introduced~\citep{Cardillo16} because this SNR is too old to explain an efficient CR acceleration. Indeed, acceleration model implies ``ad~hoc'' features, as~low-energy cut-off or broken power-law particle distribution, in~disagreement with theoretical models~\cite{Giuliani11, Cardillo14,Ackermann13}. Instead, in~the case of the $\kappa$-Ori star the $\gamma$-ray excess detected from both AGILE~\cite{Marchili18} and Fermi-LAT~\citep{Ackermann12} satellites cannot be explained neither by diffuse emission nor by contribution of ``dark gas'' (gas not traced by HI or CO molecules,~\cite{Grenier05}) ~\citep{Ackermann12, Marchili18}. This excess is correlated with a CO-detected star forming region and it could be emission due to CR energization in the interaction region between star wind FS and the CO shell. In~the very slow FS of star wind ($v_{sh}\sim$~10 km/s), DSRA is probably more efficient than acceleration~\cite{Cardillo19}.

    %****************************
    \subsection{The Supernova Remnant~W44}
    \label{Subsec:W44}
    %****************************
    
\begin{figure}[H]
\centering
 \includegraphics[scale=0.6]{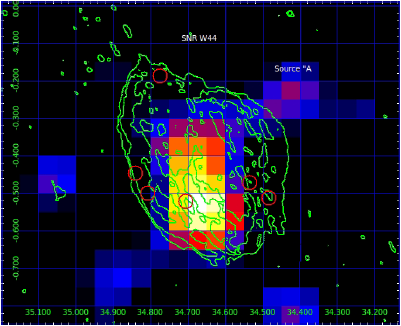}
 \includegraphics[scale=0.605]{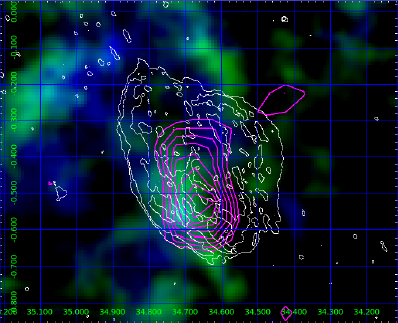}
 % W44_Cardillo14_AGILE_CO.png: 0x0 pixel, 0dpi, nanxnan cm, bb=
 \caption{The SNR W44. Left: AGILE $\gamma$-ray intensity map between 400 MeV--1 GeV of the W44 region. Green contours and red circle indicate the 324 MHz radio detection from~\citep{Castelletti07} and the OH masers from~\citep{Claussen97}, respectively.  Right: CO data from the NANTEN2 observatory. Magenta and white contours indicate the AGILE gamma-ray emission above 400 MeV and the VLA emission, respectively. Figure from~\cite{Cardillo14}.}
 %Please provide definitions id necessary.  
\label{Fig:W44_image}
\end{figure}

\textls[-10]{The SNR W44 (G34.7-0.4) is a middle-aged SNR with a distance $d\sim$~2.9 kpc~\citep{Cardillo14, Cardillo16}. Despite its location in the highly contaminated Galactic plane, it is very bright in both radio~\citep{Castelletti07} and $\gamma$-ray~\citep{Abdo10,Giuliani11,Ackermann13,Cardillo14}} wavelengths. As~stressed by presence of OH masers, its shock is interacting with a high density CO-detected MC on the SE side, in~correspondence of a peak in both the energny bands (see Figure~\ref{Fig:W44_image}).

%**********************************************************
%*****************************************************************

$\gamma$-ray emission in correspondence with high density target and of enhanced and flat radio emission could be a sign of CR acceleration at the shock. However, only after AGILE detection of the SNR W44 at energies below 200 MeV (see Section \ref{Subsec:DSRA}) energized particles  were confirmed in correspondence of the SNR/MC interaction region. One of the first acceleration models proposed a power-law particle distribution with a low-energy cut-off and a very steep high energy index ($\alpha=3$)~\citep{Giuliani11}. Some years later, the~authors of~\cite{Ackermann13,Cardillo14}, instead, suggested a broken power-law distribution, with~an energy index slightly steeper than $\alpha=2$ at low-energies and steeper than $\alpha=3$ at higher energies. A~theoretical explanation of these features was proposed in~\cite{Malkov11} where the author pointed out that Alfv'en wave damping could steepen particle spectra. However, the~slow shock velocity ($v_{sh}\sim$ 100--150 km/s) due to the high age of W44 questioned the presence of freshly accelerated~particles. 

After the first attempts to model radio and $\gamma$-ray spectra of W44 using ``ad~hoc'' spectral features or neglected leptonic emission and energy losses~\citep{Uchiyama10,Lee15,Tang15}, new Voyager I data raised the opportunity to develop a more realistic re-acceleration model (see Section~\ref{Subsec:Voyager}). In~\cite{Cardillo16}, these data were used to compute, analytically and numerically, $\gamma$-ray emission from the SNR W44 assuming a DSRA with crushed adiabatic shell formation, according to the model explained in Section~\ref{Sec:model}. Assuming a Kolmogorov-like turbulence ($k_{T}=5/3$) and considering the diffusion coefficient dependences~\citep{Ptuskin03,Cardillo16}, the~expression for maximum momentum is derived from Equation~(\ref{Eq:pmax_gen}):
%---------------------------------------------------
\begin{equation}
p_{\rm max}\sim7\,GeV/c\,\left(\frac{B_{0}}{30\,\mu G}\right)\left(\frac{v_{sh}}{130 \,km/s}\right)^{6}\left(\frac{min(t_{int},t_{loss})}{15000\,years}\right)^{3}\left(\frac{L_{c}}{0.1\,pc}\right)^{-2}\ ,
\label{Eq:EmaxKo_W44}
\end{equation}
%------------------------------------------------------
This formula clearly shows that shock velocity is the most critical parameter and, consequently, explains the low-energy value of the cut-off. In~this model a filling factor $\xi$ taking into account the SNR evolution was used,  $V=4\pi\xi R^{2}_{SN}v_{sh}t_{int}$.

\begin{figure}[H]
\centering
\includegraphics[scale=0.6]{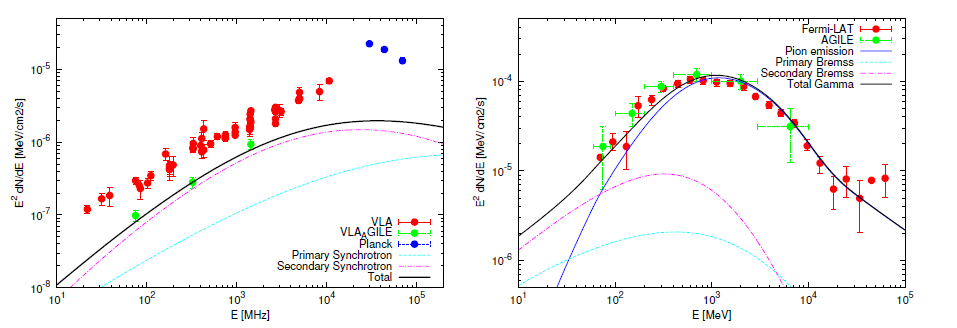} 
\caption{Radio and $\gamma$-ray data from W44 fitted with the best re-acceleration model. Left: VLA (red) and Planck (blue) radio data from the whole remnant and VLA radio data from the AGILE emitting region (green). Primary, secondary, and~total synchrotron radio emission from the model are indicated from cyan dashed line, magenta dot-dashed and black line, respectively. Right: AGILE (green) and Fermi-LAT
(red) $\gamma$-ray points plotted with $\gamma$-ray emission from pion decay (blue dotted line), from~bremsstrahlung of primary (cyan dashed line), and~secondary (magenta dot-dashed line) electrons, and~total emission (black line). Figure from~\citep{Cardillo16}.}
\label{Fig:W44_spectrum}
\end{figure}

The best model fitting both radio and $\gamma$-ray spectra (shown in Figure~\ref{Fig:W44_spectrum}) considers re-acceleration as the main contributor, with~a smaller contribution from acceleration ($\xi_{CR}\sim$~10$^{-4}$) necessary to reduce the value of the filling factor compared to the one obtained using a model with only re-acceleration ($\xi\sim0.55$) \cite{Cardillo16}.

%******************************* Figure final spectra reacceleration
%***********************************************************

This model, however, has some critical aspects. Such a low maximum energy does not constrain momentum spectral index $\alpha$, that can vary in a range $[4.0\div4.2]$ without significant changes in the main conclusions. This implies that real spectral index could be steeper than 4, as~in all the other SNRs detected in the $\gamma$-ray band, and~we do not understand the reason yet. Moreover, this model was developed assuming a totally ionized medium at the shock. If~this assumption was not correct, there could be ion-neutral damping inhibiting re-acceleration~\citep{Drury96,Ptuskin03}. For~this reason, in~\cite{Cardillo16} the authors showed a model that considers only crushed cloud adiabatic compression without any first order Fermi energization. This model can still explain W44 $\gamma$-ray emission but with a worst~fit.

Nevertheless, the~SNR W44 remains the first case in which a DSRA model account for a source multi-wavelength spectrum in agreement with theory and with physically coherent~parameters.  

     %****************************
     \subsection{The $\kappa$-Ori~Wind }
     \label{Subsec:Orion}
     %****************************

The Orion region is very interesting because it is the nearest site of star formation ($d\sim$~400--450~pc) and it is unaffected by Galactic diffuse emission~\cite{Bally08}. It includes different structures: molecular clouds, about one hundred of OB stars and several young stellar objects (YSOs), as~revealed by the X-ray satellite XMM-Newton in the last years. These objects are in a high density million years old shell along the line of sight of the MC Orion A~\cite{Pillitteri16} that seems to be correlated with $\kappa$-Ori star. It could have triggered star formation in the shell through its~wind.

\begin{figure}[H]
\centering
 \includegraphics[scale=0.75]{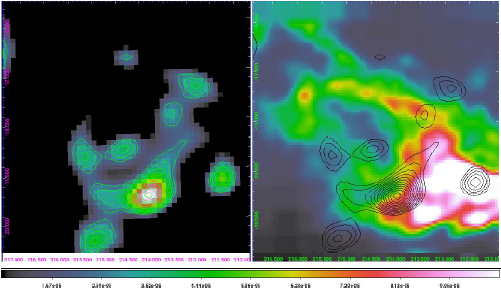}
 % AGILE_shell.png: 0x0 pixel, 0dpi, nanxnan cm, bb=
\caption{The Orion region around $\kappa$-Ori OB star. Left: $\gamma$-ray excess detected by AGILE. Right panel: CO map from~\citep{Dame01} with contour levels from $\gamma$-ray data in black. The~figure is from~\citep{Marchili18}.}
\label{Fig:Orion_AGILE}
\end{figure}

COS-B~\cite{Caraveo80} and EGRET~\cite{Digel95} satellites did the first $\gamma$-ray surveys of this region; however, only recent observations from Fermi-LAT~\cite{Ackermann12} and AGILE~\cite{Marchili18} revealed an interesting feature. Data from the two satellites characterized $\gamma$-ray diffuse emission from the Orion region, taking into account Bremsstrahlung and p-p emission from atomic and molecular Hydrogen, inverse Compton from InterStellar Radiation Field (ISRF) and CMB, and~single source contribution. The~$\gamma$-ray residual map shows an excess at the South-West side of $\kappa$-Ori, in~correspondence of the high-density shell detected by XMM-Newton (see Figure~\ref{Fig:Orion_AGILE}).

%**************************************************
%***************************************************

A first idea to explain this excess was to add ``dark gas'' \cite{Grenier05} contribution to diffusion model. This was computed through column density derived from reddening maps of IRAS and COBE and from Planck 353 GHz maps. The~residuals of the subtraction of CO-traced MC contributions are used as ``dark gas'' templates~\cite{Marchili18}. However, both Fermi-LAT and AGILE analysis pointed out that ``dark gas'' has only a little effect on the $\gamma$-ray excess flux. Another explanation could be that a possible non-linearity in CO-$H_{2}$ relation affects the constance of the $X_{CO}$ conversion factor. This parameter is fundamental in $\gamma$-ray diffuse models because it allows to estimate $H_{2}$ density from $CO$ detection~\citep{Ackermann12, Marchili18}. The~recent AGILE analysis of this excess revealed its relation with the high density shell of~\cite{Pillitteri16} and found a spectrum with a hard index that could explain its origin in the context of CR energization~\citep{Marchili18, Cardillo19}. 

Stellar wind shocks are considered good locations for CR acceleration~\citep{Casse80, Voelk82, Cesarsky83, Ip95} but only the Termination Shocks (TS) of star winds were taken into account so far, because~of their high velocity ($10^{2}$--$10^{3}$ km/s). In~the model described in~\cite{Cardillo19}, instead, CR energization takes place in correspondence of the slow FS  (order of $10$ km/s) of $\kappa$-Ori star, assuming that the $\gamma$-ray excess could be due to pre-existing CR~re-acceleration. 

The authors used the model described in Section~\ref{Sec:model}, fixing some important parameters thanks to information from other wavelengths, such as distance, $d\sim$~{240}--{280} pc, and~age, $t_{age}\sim$~7~$\times$~10$^{7}$~years, of~$\kappa$-Ori and FS velocity. The~average density in the region of the shell was estimated through the conversion factor $X_{CO}$ computed in~\cite{Marchili18} and it is equal to $n_{0}\sim$~30 cm$^{-3}$. Lack of radio and TeV detection allowed us to constraint other parameters such as magnetic field and perturbation correlation length~\citep{Cardillo19}. We used the conservative Kolmogorov spectrum also in this case, providing $\delta=k_T-1=\frac{2}{3}$. In~this way, we obtain an explicit expression for maximum momentum:
%------------------------------- Maximum momentum Kolmogorov
%\begingroup\makeatletter\def\f@size{8}\check@mathfonts
\begin{equation}
p_{\rm max}\sim41.8\,GeV/c\,\left(\frac{B_{0}}{15\,\mu G}\right)\left(\frac{v_{sh}}{10 \,km/s}\right)^{6}\left(\frac{t_{min}}{700000\,years}\right)^{3}\left(\frac{L_{c}}{0.01\,pc}\right)^{-2}\
\label{Eq:EmaxKo_Orion}
,\end{equation}
%\endgroup
%----------------------------------------------
where normalization values are of the same order assumed (or estimated) in our best model. Looking at the numerical value of this equation, it is clear that we expect a cut-off at low energies, excluding the possibility of TeV emission from this~region.

\begin{figure}[H]
\centering
 \includegraphics[scale=0.45]{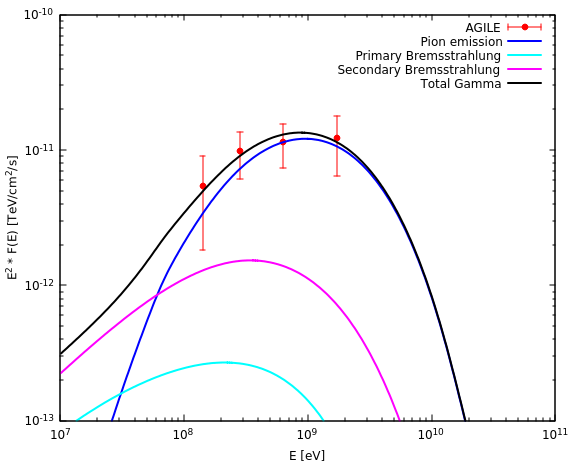}
 % Gamma_Modellov11_2R_paper.png: 0x0 pixel, 0dpi, nanxnan cm, bb=
 \includegraphics[scale=0.45]{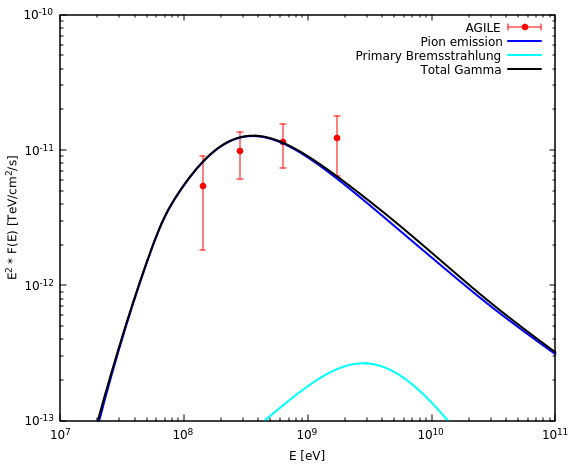}
 % Gamma_Modello_compression_only_v11_2c_paper.png: 0x0 pixel, 0dpi, nanxnan cm, bb=
\caption{Left: AGILE (red) $\gamma$-ray points plotted with the different contributions estimated in our re-acceleration model: $\gamma$-ray emission from pion decay (blue line), primary and secondary Bremsstrahlung (cyan dashed line and magenta dashed line, respectively), and~total emission (black line). Right: the same data and model curves are shown in the left panel but assuming that only adiabatic compression is present. Figures from~\citep{Cardillo19}}
\label{Fig:Orion_model}
\end{figure}

DSRA can explain the AGILE detected $\gamma$-ray emission, regardless of presence of an adiabatic compressed shell, because~energy losses compensate for the enhancement due to higher densities (Figure \ref{Fig:Orion_model}, left). However, still considering a possible presence of neutrals at the shock, the~authors have taken into account the chance to have CR energization due to the only adiabatic compression and found that it could be sufficient to explain the AGILE emission as well (Figure \ref{Fig:Orion_model}, right). 

%********************************************
%*********************************************************

Unfortunately, because~of the cut-off at such a low-energy, as in the SNR W44, there are no strong constraints for momentum spectral index that can vary in a range [5 -- 3.5]. Nevertheless, some important issues can be fixed: the model does not provide for radio or TeV emission, in~agreement with absence of detections in those energy bands, and~freshly accelerated particle emission can be excluded as the dominant~contribution. %Please check if this should be a multiplication

%%%%%%%%%%%%%%%%%%%%%%%%%%%%%%%%%%%%%%%%%%
\section{Conclusions}
\label{Sec:Conclusions}
%%%%%%%%%%%%%%%%%%%%%%%%%%%%%%%%%%%%%%%%%%

CR origin is one of the most important issues of high energy astrophysics of the last decades. Looking for CR sources, we gained a better understanding of physical mechanisms correlated to magnetic turbulence and perturbations, and~we could deeply analyze several objects, such as SNRs, extracting their peculiar spectral~features.

Thanks to the great improvement of high energy instrument performances, we collected a large amount of data that have questioned theoretical models, pointing out the complexity of CR behavior. In~this brief review we summarized the significant progress made introducing CR re-acceleration to explain CR particle and $\gamma$-ray~data.

We verified that stochastic and diffusive shock re-acceleration can contribute in a non-negligible way to the explanation of Boron over Carbon ratio, CR spectral hardening and anti-proton excess detected by PAMELA and AMS in particle spectrum. In particular, we showed that DSRA could be the dominant energization process in middle-aged SNRs like W44, where the high age makes particle acceleration inefficient. DSRA and adiabatic compression due to thin dense shell formation can explain both AGILE and Fermi-LAT spectra according to theoretical models, without~introduction of any ``ad~hoc'' feature. Moreover, DSRA could be efficient in other kinds of objects, such as stellar wind. We described the case of $\kappa$-Ori stellar wind in the Orion region, where a $\gamma$-ray excess detected by Fermi-LAT and AGILE could be explained by CR re-acceleration at the interaction region with a CO-detected star formation shell. Whereas freshly accelerated particles at TS of a stellar wind were already predicted, this is the first time that high energy energized particle emission is detected at the slow FS of a~star.

The important contribution of re-acceleration mechanism does not mean that it is the dominant process in all sources analyzed so far or that it could explain every particular feature of CR spectrum. Fresh particle acceleration remains the main CR production process, even if we do not have direct evidence yet. However, we want to highlight that, before~introducing new physical mechanisms and challenging our theoretical models, we have to consider all known physical processes in order to understand CR origin, as is the case~for any other tricky scientific~issue.

\vspace{6pt} 

%%%%%%%%%%%%%%%%%%%%%%%%%%%%%%%%%%%%%%%%%%
\funding{This research received no external~funding.}

%%%%%%%%%%%%%%%%%%%%%%%%%%%%%%%%%%%%%%%%%%
%\acknowledgments{}

%%%%%%%%%%%%%%%%%%%%%%%%%%%%%%%%%%%%%%%%%%
\conflictsofinterest{The author declares no conflict of~interest.} 

%=====================================
% References, variant A: internal bibliography
%=====================================
\reftitle{References}

% The following MDPI journals use author-date citation: Arts, Econometrics, Economies, Genealogy, Humanities, IJFS, JRFM, Laws, Religions, Risks, Social Sciences. For those journals, please follow the formatting guidelines on http://www.mdpi.com/authors/references
% To cite two works by the same author: \citeauthor{ref-journal-1a} (\citeyear{ref-journal-1a}, \citeyear{ref-journal-1b}). This produces: Whittaker (1967, 1975)
% To cite two works by the same author with specific pages: \citeauthor{ref-journal-3a} (\citeyear{ref-journal-3a}, p. 328; \citeyear{ref-journal-3b}, p.475). This produces: Wong (1999, p. 328; 2000, p. 475)

%=====================================
% References, variant B: external bibliography
%=====================================
%\externalbibliography{yes}
%\bibliography{your_external_BibTeX_file}

%%%%%%%%%%%%%%%%%%%%%%%%%%%%%%%%%%%%%%%%%%
%% optional
%\sampleavailability{Samples of the compounds ...... are available from the authors.}

%% for journal Sci
%\reviewreports{\\
%Reviewer 1 comments and authors’ response\\
%Reviewer 2 comments and authors’ response\\
%Reviewer 3 comments and authors’ response
%}

%%%%%%%%%%%%%%%%%%%%%%%%%%%%%%%%%%%%%%%%%%
\end{document}